\documentclass[prb,aps,twocolumn,showpacs,floatfix,superscriptaddress]{revtex4}
\usepackage{graphicx}
\usepackage{amsfonts}
\usepackage{amssymb}
\usepackage{amsmath}
\usepackage{enumerate}
\usepackage{multirow}
\usepackage{color}
\usepackage{subfig,float}
\usepackage{empheq}
\newcommand{\Hamil}{{\cal H}}

\newcommand{\be}{\begin{equation}}
\newcommand{\ee}{\end{equation}}
\newcommand{\ba}{\begin{eqnarray}}
\newcommand{\ea}{\end{eqnarray}}
\newcommand{\up}{\uparrow}
\newcommand{\down}{\downarrow}

\newcommand{\mb}[1]{\mathbf{#1}}

\begin{document}


\title{Spin and Majorana polarization in topological superconducting wires }
\author{Doru Sticlet}
\affiliation{Laboratoire de Physique des Solides, CNRS UMR-8502, Univ. Paris Sud, 91405 Orsay Cedex, France}
\author{Cristina Bena}
\affiliation{Laboratoire de Physique des Solides, CNRS UMR-8502, Univ. Paris Sud, 91405 Orsay Cedex, France}
\affiliation{Institute de Physique Th\a'eorique, CEA/Saclay, Orme des Merisiers, 91190 Gif-sur-Yvette Cedex, France}
\author{Pascal Simon}
\affiliation{Laboratoire de Physique des Solides, CNRS UMR-8502, Univ. Paris Sud, 91405 Orsay Cedex, France}

\date{\today}

\begin{abstract}
We study a one-dimensional wire with strong Rashba and Dresselhaus spin-orbit coupling (SOC), which supports Majorana fermions when subject to a Zeeman magnetic field and in proximity of a superconductor. Using both analytical and numerical techniques we calculate the electronic spin texture of the Majorana end states. We find that the spin polarization of these states depends on the relative magnitude of the Rashba and Dresselhaus SOC components. Moreover, we define and calculate a local ``Majorana polarization'' and ``Majorana density'' and argue that they can be used as order parameters to characterize the topological transition between the trivial system and the system exhibiting Majorana bound modes.
We find that the local ``Majorana polarization'' is correlated to the transverse spin polarization, and
we propose to test the presence of Majorana fermions in a 1D system by a spin-polarized density of states measurement.
\end{abstract}

\pacs{73.20.-r, 74.78.Fk, 73.63.Nm}
\maketitle

{\it Introduction}--Majorana fermions have been attracting a lot of interest recently in the light of the discovery of new materials with  topological-insulator properties \cite{topological}. These atypical fermionic particles
have been predicted long time ago by E. Majorana as real solutions of the Dirac equation \cite{majorana}.
Many condensed-matter systems such as Pfaffian states in fractional quantum
Hall (FQH) systems \cite{read91},  chiral $p$-wave superconductors \cite{ivanov01} (like strontium ruthenate\cite{maeno03}), nodal superconductors under certain conditions \cite{sato10}, ultracold fermionic atoms with laser-field-generated spin-orbit intearctions \cite{sato09},
the surface of  3D strong topological insulators \cite{fu08}, as
well as semiconductor/superconductor heterostructures \cite{hetero}
have been proposed as platforms supporting Majorana fermions.
Among the possible heterostructures,  one-dimensional (1D) systems with a strong spin-orbit coupling such as  InAs and InSb  wires \cite{inas}, subject to Zeeman magnetic field and in proximity of a superconductor (SC), can  exhibit Majorana fermions at their extremities \cite{lutchyn10,oreg10}.
 Various proposals have been made to detect the Majorana states including interferometry \cite{transport}, noise \cite{noise} and spectroscopy measurements \cite{spectroscopy}, etc. However, while a direct confirmation of the existence of the Majorana states would constitute an important step for fields such as quantum computation \cite{qc}, they have not been detected experimentally so far.

Majorana modes for two dimensional spin-triplet topological superconductors have been shown to exhibit an Ising-like spin density that may allow their detection via coupling to a magnetic impurity\cite{shindou}. Along similar lines, we propose a method to detect the Majorana states in 1D topological semiconducting wire spectroscopically, using spin-polarized scanning tunneling microscopy (STM). We generalize the model in Refs.~\onlinecite{oreg10,lutchyn10} to include both Rashba and Dresselhaus spin-orbit interactions, and we show that the resulting Majorana bound states exhibit a characteristic spin texture. In particular we find that the  component of the spin polarization in the transverse spin-plane (orthogonal to the direction of the magnetic field) is non-zero solely in the topological phase, and  its orientation is determined by the relative weight of the SOC components.  Moreover, we introduce a ``Majorana pseudospin'' local order parameter and define two new local quantities denoted Majorana polarization and Majorana density which quantify locally the Majorana character of a state.  We show that the transverse spin polarization is related to the Majorana polarization and we propose that its measurement via spin-polarized STM will allow one to directly visualize the Majorana fermionic states, and thus to test the topological character of a one-dimensional system.

{\em Model}-- We consider a semiconducting wire oriented along the $x$-direction, and in proximity to a $s$-wave superconductor. Due to bulk inversion asymmetry, semiconducting wires can exhibit along with the Rashba SO interaction analyzed in Refs.~\onlinecite{oreg10,alicea11}, a Dresselhaus SO interaction of the same order of magnitude\cite{loss} ($\sim$ 0.1 eV \AA).
The Bogoliubov-de Gennes (BdG) Hamiltonian for the infinite wire with both types of SOC can be written as
\begin{align}\label{ham}
H&=\int\Psi^\dag\Hamil\Psi dx,\quad
\Psi^\dag=(\psi^\dag_\uparrow,\psi^\dag_\downarrow,\psi_\downarrow,-\psi_\uparrow),\notag\\
\Hamil&=\bigg(\frac{p^2}{2m}-\mu+\alpha p\sigma_y+\beta p\sigma_x\bigg)\tau_z+
V_z\sigma_z-\Delta\tau_x.
\end{align}
$\sigma$'s and $\tau$'s are the usual Pauli matrices acting respectively in the spin and particle-hole spaces. The chemical potential is denoted by $\mu$, $V_z$ is the Zeeman field, $\Delta$ is the induced superconducting pairing and $\alpha$ ($\beta$) characterize the strengths of the Rashba (Dresselhaus) SOC components. The presence of the Dresselhaus term only trivially modifies the spectrum for the translationally invariant system\cite{oreg10}
\begin{align}
E^2&=\xi^2+(\alpha^2+\beta^2)p^2+V_z^2+\Delta^2&\notag\\
&\quad\pm 2(\xi^2(\alpha^2+\beta^2)p^2+\xi^2V_z^2+\Delta^2V_z^2)^{1/2},
\end{align}
with $\xi=p^2/2m-\mu$.
A careful analysis of this model shows that the conditions for the existence of the topological phase supporting Majorana fermions are unaffected by the Dresselhaus SO interactions, $V_z^2>\Delta^2+\mu^2$. It is interesting to note however that Majorana bound states can exist even in the absence of the Rashba term, when  Dresselhaus SO interactions are present. Most importantly, the spin texture of the Majorana states is influenced by the presence of the Dresselhaus term. To support this claim we present an analytical study of the wavefunctions corresponding to the Majorana bound states, and we complement it by a numerical study of the corresponding lattice model.

{\em Analytical solution}--It has been shown that Majorana bound states can arise at the interface between trivial and topological regions of a one dimensional wire, for example by considering a position dependent chemical potential \cite{oreg10,gil}. Thus, by choosing a chemical potential $\mu_1$ for $x\in[0,L]$, such that $\mu_1^2<V_z^2-\Delta^2$, and a $\mu_0^2>V_z^2-\Delta^2$ outside this interval, one obtains a finite-size topological region inside a topologically trivial phase. The chemical potentials are chosen such that the $p=0$ gap, $\Delta-\sqrt{V^2-\mu^2}$, is much smaller than the superconducting gap $\Delta$, which allows one to obtain analytical solutions to the problem by linearizing the Hamiltonian in $p$.
We assume that $L\gg 1$, such that the problem can be solved independently at the two ends. Thus, ignoring the finite-size effects, the condition to have zero-energy solutions bound at the two interfaces yields the allowed values for the momenta, $\sqrt{\alpha^2+\beta^2}k_j^{\pm}=\Delta\pm\sqrt{V^2_z-\mu_j^2}$, $j\in\{0,1\}$.
Consequently, the Majorana solution at the left boundary $\psi(x\sim 0)$ can be written as
\begin{empheq}[left={\empheqlbrace}]{align}
&\kappa\mb u_1(\mu_1)e^{k^{-}_1x},& x>0,\notag\\
&\frac{\kappa}{2}\bigg[(1+\frac{\tan\phi_1}{\tan\phi_0})\mb u_1(\mu_0)e^{k^{-}_0x}\\
&+(1-\frac{\tan\phi_1}{\tan\phi_0})\mb u_2(\mu_0)e^{k^{+}_0x}\notag
\bigg],
&x<0.
\end{empheq}
Similarly, the Majorana solution at the right boundary $\psi(x\sim L)$ is
\begin{empheq}[left={\empheqlbrace}]{align}
&\kappa\mb u_3(\mu_1)e^{-k^{-}_1(x-L)},&x<L\notag\\
&\frac{\kappa}{2}\bigg[(1+\frac{\tan\phi_1}{\tan\phi_0})
\mb u_3(\mu_0)e^{-k^{-}_0(x-L)}\\
&+(1-\frac{\tan\phi_1}{\tan\phi_0})\mb u_4(\mu_0)e^{-k^{+}_0(x-L)}\notag
\bigg],&x>L
\end{empheq}
with
$e^{i\phi_j}=1/\sqrt{2}(\sqrt{1+\mu_j/V_z}+i\sqrt{1-\mu_j/V_z})$, and the Majorana eigenvectors are given by
\begin{align}
\mb u_1(\mu_j)^T&=(\cos\phi_je^{i\vartheta},-\sin\phi_j,\sin\phi_je^{i\vartheta},\cos\phi_j),\notag\\
\mb
u_2(\mu_j)^T&=(\cos\phi_je^{i\vartheta},\sin\phi_j,-\sin\phi_je^{i\vartheta},\cos\phi_j),\notag\\
\mb
u_3(\mu_j)^T&=-(\cos\phi_je^{i\vartheta},\sin\phi_j,\sin\phi_je^{i\vartheta},-\cos\phi_j),\\
\mb
u_4(\mu_j)^T&=(-\cos\phi_je^{i\vartheta},\sin\phi_j,\sin\phi_je^{i\vartheta},\cos\phi_j)\notag.
\end{align}
Here $e^{i\vartheta}=(\alpha+i\beta)/\sqrt{\alpha^2+\beta^2}$ allows one to define a two-dimensional spin-orbit vector in the transverse plane, ${\bf e}_{\vartheta}=(\cos\vartheta,\sin\vartheta)$.
Note that the obtained wavefunctions are indeed Majorana fermions respecting the reality condition through the phase choice $(\vartheta+\pi)/2$ for the complex coefficient $\kappa$. The magnitude of $\kappa$ is determined from the normalization conditions of the wave functions and is of the order of $(\sqrt{V_z^2-\mu_1^2}-\Delta/\sqrt{\alpha^2+\beta^2})^{1/2}$.

The electronic local spin polarization $\mb s(x)$ of a given four-component (two spin$\times$ electron/hole) Nambu state $|\psi\rangle$ can be calculated by evaluating the expectation values $s_i(x)=\langle \psi|\sigma_i\frac{\tau_0+\tau_z}{2}|\psi\rangle$ (the $\tau_0+\tau_z$ insures that we only take into account the electronic, and not the hole degrees of freedom).
This prescription yields for the Majorana fermionic states at $x=0$ and $x=L$:
\begin{align}\label{spin}
\mb s(0)
&=\frac{|\kappa|^2}{2}(-\sin(2\phi_1)\cos\vartheta,\sin(2\phi_1)\sin\vartheta,\cos(2\phi_1))\notag\\
\mb s(L)
&=\frac{|\kappa|^2}{2}(\sin(2\phi_1)\cos\vartheta,-\sin(2\phi_1)\sin\vartheta,\cos(2\phi_1)).
\end{align}
The above results show that the spin $z$-components are equal at the ends of the wire, while the transverse spin polarization is equal in magnitude and opposite. Its direction is fixed solely by the relative weight of the Rashba and Dresselhaus SOC components, $s_y/s_x=-\beta/\alpha$. It would be interesting to see whether this results
also holds in presence of interactions \cite{alicea11,simon11}.


{\em Definitions}--A general wave function written in Nambu basis described in Eq. (\ref{ham}) as $(u_\up,u_\down,v_\down,v_\up)$ can be recast
in a Majorana basis  $(\gamma_{1\up},\gamma_{2\up},\gamma_{1\down},\gamma_{2\down})$,  where
\begin{equation}
(\gamma_{1\delta},\gamma_{2\delta})=\frac{1}{\sqrt{2}}
(\psi_\delta^\dag e^{-i\varphi/2}+\psi_\delta e^{i\varphi/2},
i(\psi_\delta^\dag e^{-i\varphi/2}-\psi_\delta e^{i\varphi/2})),
\end{equation}
and $\varphi$ is a phase characterizing the particular choice of  Majorana basis.
We can focus simply on the spin up part of the wave function and define the Majorana polarization $P_M^\up$ as the difference between the probability to have a $\gamma_{1\up}$ Majorana and the probability to have a $\gamma_{2\up}$ Majorana,
$P_M^\up=P_{\gamma_1\up}-P_{\gamma_2\up}$. Note that $P_{\gamma_{1/2}}$  but not $P_M$, have been also introduced in Ref. \onlinecite{alicea11}. For the generalized ($\varphi$ not fixed) Majorana basis, it reads $P_M^\up=-2{\rm Re}[u_\up v^*_\up e^{-i\varphi}]$. The phase $\varphi$ defines a ``Majorana polarization axis'' such that $P_M^\up$ can be interpreted as a vector (we can denote this vector as ``Majorana pseudopsin''). As such it is decomposed on the axis defined by $\varphi$ into $P^\up_{M_x}(\varphi=0)=-2{\rm Re}[u_\up v^*_\up]$ and $P^\up_{M_y}(\varphi=\pi/2)=-2{\rm Im}[u_\up v^*_\up]$.

The same procedure can be repeated for the spin down component. This leads to the definition of the full Majorana polarization,  $P_{M_i}=P^\up_{M_i}+P^\down_{M_i}$ with $i\in\{x,y\}$. We should note that the absolute value of the polarization vector $\mathcal P_M=\sqrt{P_{M_x}^2+P_{M_y}^2}$, which we denote Majorana density, is not dependent on the choice of ($x$,$y$) axes in the Majorana space.
For further reference the {\em Majorana polarization} and {\em density} are
\begin{align}\label{pol}
P_{M_x}&=2{\rm Re}[u_\down v^*_\down-u_\up v^*_\up],&
P_{M_y}&=2{\rm Im}[u_\down v^*_\down-u_\up v^*_\up],\notag\\
\mathcal P_M&=2|u_\down v^*_\down-u_\up v^*_\up|.
\end{align}

To emphasize the utility of such definitions the following properties should be noted. For a wave function containing only electronic (or hole) degrees of freedom, the Majorana polarization is always zero. It equally vanishes for a conventional $s$-wave superconductor where the wave function has the symmetry $u^{\phantom*}_{(\down,\up)}=v^*_{(\up,\down)}$. A non-zero value for the Majorana density is a necessary, but not sufficient condition to have Majorana fermions, and identifies the low energy regime in which the model yields an effective $p$-wave type superconductivity. At zero energy this quantity is maximal and predicts Majorana bound states at the edges. Inspecting the Majorana polarization of these states allows one to unambiguously identify  that the two Majorana fermions are of different type. The most important property is that such definitions allow one to explore locally (on-site in the discretized system) the structure of the wave function, and its Majorana character.

For the above analytical solutions, the Majorana polarization vectors $\mb P_M=(P_{M_x},P_{M_y})$ yield
\begin{equation}\label{majpol}
\mb P_M(0)=-\mb P_M(L)=-|\kappa|^2(\cos\vartheta,\sin\vartheta\cos(2\phi_1))
\end{equation}
The Majorana polarization vectors are opposite for the two Majorana wave functions at the two ends. Besides, one can identify a relation between the spin polarization vector and the Majorana polarization. When $\mu$, $\Delta$, $V_z$ are fixed,
$P_{M_x}$ is proportional to $s_x$, while $P_{M_y}$ is proportional to $ s_y$. Thus, when only Rashba/Dresselhaus SOC is present, the total transverse spin polarization is proportional to  the Majorana polarization, with a proportionality constant which depends on the chemical potential potential and the applied Zeeman field. When both components of the SOC are present, the Majorana polarization  and the transverse spin polarization vectors are no longer collinear but the correlations between these two quantities  are still qualitatively preserved.

{\em Numerical model}--For the numerical study, we consider a tight-binding formulation of BdG Hamiltonian in Eq. (\ref{ham})
\begin{align}\label{hdis}
H&=\sum_j\Psi_j^\dag[(\mu-t)\tau_z+V_z\sigma_z-\Delta\tau_x]\Psi_j\notag\\
&\quad-\frac{1}{2}\bigg[\Psi_j^\dag(t+i\alpha\sigma_y+i\beta\sigma_x)\tau_z\Psi_{j+1}
+{\rm h.c.}
\bigg]
\end{align}
with the Nambu basis $\Psi^\dag_j=(c^\dag_{j\up},c^\dag_{j\down},c_{j\down},-c_{j\up})$. The sum is performed over $N=100$ sites in the system.
We work in units of $t=1$ and we consider the lattice constant $l=1$ and $\hbar=1$. Numerical simulations are done for typical values of the parameters $V_z=0.4,\Delta=0.3,\mu=0$.

By exact diagonalization we have access to the local density of states, and the local spin-polarized density of states along the $x$, $y$, and $z$ directions. For example the local (site $n$) electronic $i$-spin polarization density is given by
\begin{equation}
\rho_{{\sigma_i},n}=\sum_{j=1}^{4N}\delta(\omega-E_j)
\langle\Psi^{(j)}_n|\sigma_i\frac{\tau_0+\tau_z}{2}|\Psi^{(j)}_n\rangle
\end{equation}
where $E_j$ is the $j^{\rm th}$ eigenvalue of $H$ and $|\Psi_n^{(j)}\rangle$ is the site $n$ component of the $j^{\rm th}$ eigenvector, $|\Psi_n^{(j)}\rangle=(u^{(j)}_{n\up},u^{(j)}_{n\down},v^{(j)}_{n\down},v^{(j)}_{n\up})$.
We can thus use the definitions in Eq.~(\ref{pol}) to calculate the local Majorana polarization and density. For example, the local ($n$ site) Majorana $x$-polarization can be written as
\begin{equation}
P_{M_x,n}=\sum_{j=1}^{4N}\delta(\omega-E_j)
2{\rm Re}[u^{(j)}_{n\down}v^{*(j)}_{n\down}-u^{(j)}_{n\up}v^{*(j)}_{n\up}].
\end{equation}
Numerically, we implement the $\delta$ functions as very thin Gaussians of width $\approx 0.0001 \hbar v_F/l$.

\begin{figure}[ht]
\includegraphics[width=.4\textwidth]{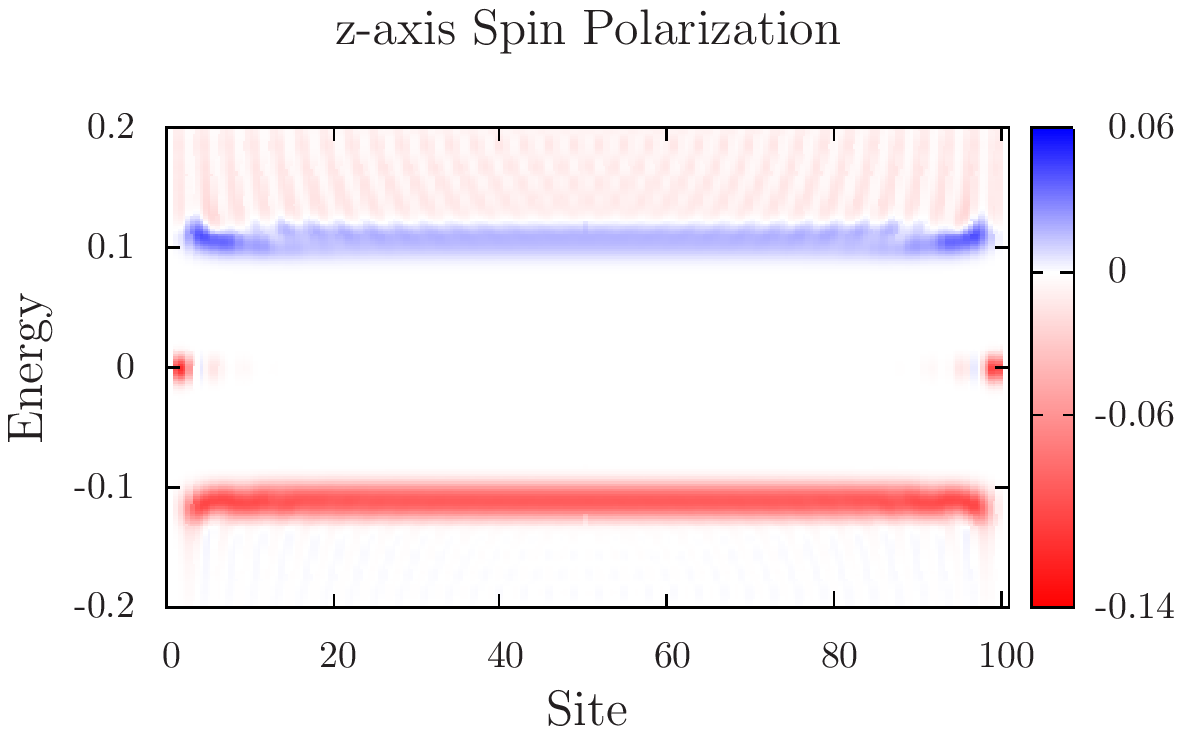}
\vspace{-0.1in}\\
\includegraphics[width=.4\textwidth]{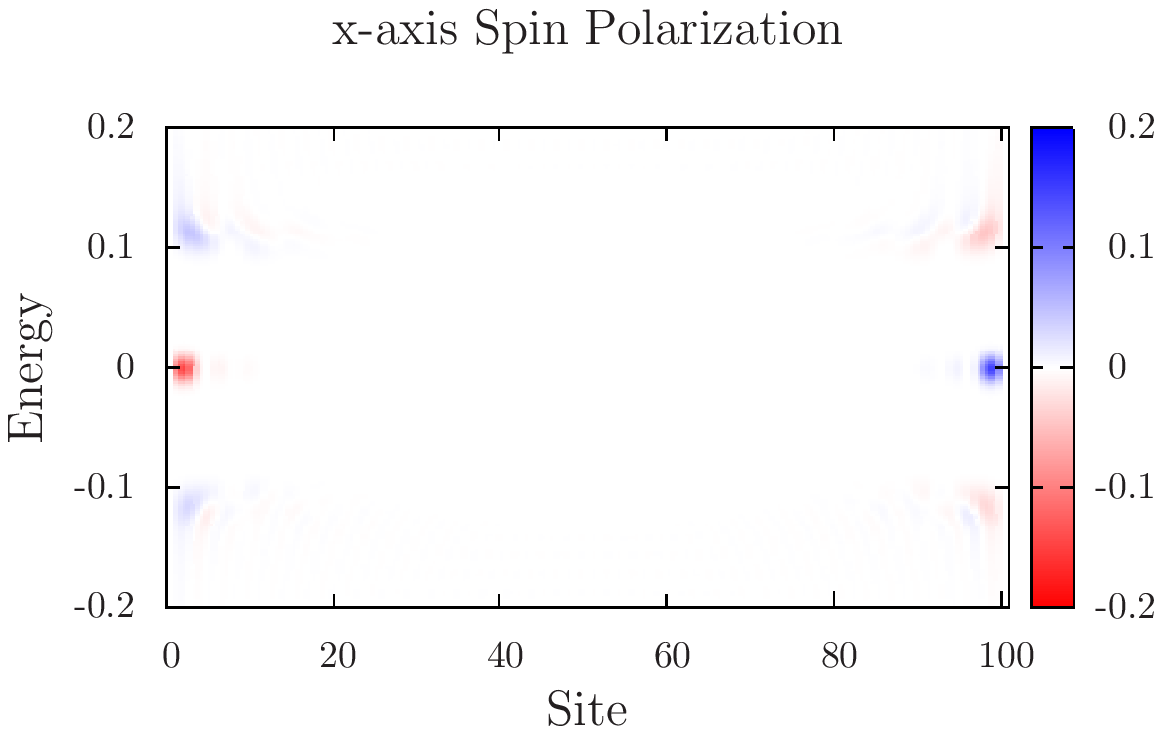}
\vspace{-0.1in}\\
\includegraphics[width=.4\textwidth]{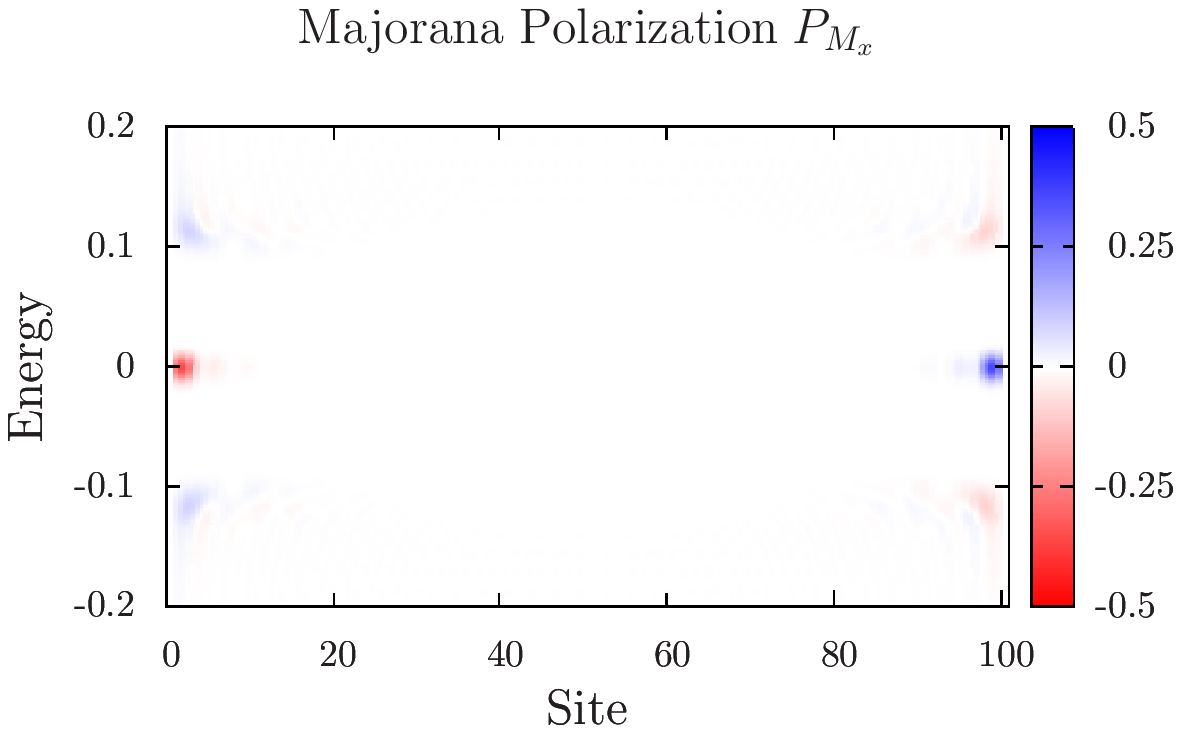}
\vspace{-0.1in}
\caption{\small The spin polarization along the $z$ and $x$ directions, and the Majorana polarization $P_{M_x}$, as a function of energy and position, $\Delta=0.3$, $V_z=0.4$, $\alpha=0.2$, $\beta=0$ and $\mu=0$.}
\label{fig:xz}
\end{figure}

{\em Numerical results}--As expected, the exact diagonalization of $H$ recovers the Majorana bound states at the ends of the wire.
In Fig. (\ref{fig:xz}) we present the $x$ and $z$ components of the spin polarization, as well as the Majorana polarization when the only considered SOC is Rashba. The symmetrical situation, with only the Dresselhaus component of the SOC  present, is analyzed in the Supplementary Material (SM). We can see that the zero-energy Majorana wavefunctions are extended over a small number of edge sites, while exhibiting strongly damped spatial oscillations.
The transverse electronic spin polarization is opposite at the two ends of the wire; when only the Rashba term is present only the $x$-component of the transverse spin is non-zero. The $z$-spin polarization is the same at the two ends of the wire. While
Eq. (\ref{spin}) predicts that
 the $z$-polarization vanishes for $\mu_1=0$, this is not the case in the numerical calculation. This can be understood by the presence of an effective $\mu$ due to the neglected kinetic term, which to leading order, contributes to $\langle p^2\rangle \approx O((\Delta-V_z)^2)$, which creates a {\em negative} effective potential as in the numerical results.
Moreover, this effective chemical potential is responsible for the spatial (quickly damped) oscillations of the spin polarization observed numerically. Allthough these oscillations are not captured by the continuum limit calculations, one can however
check that the ratio $s_{y,i}/s_{x,i}$ depends only on the spin-orbit couplings in agreement with Eq. (\ref{spin}).


The numerical results for the Majorana polarization presented in Fig. (\ref{fig:xz}) also follow closely Eq.~(\ref{majpol}). Thus, the values of the Majorana polarization are always opposite at the two ends of the wire, and only $P_{M_x}$ is non-zero when only the Rashba component is present. Also, as expected, $P_{M_x}$ is proportional in this case to the $x$-spin polarization. Moreover, while we do not present this here in detail, when both the Rashba and Dresselhaus SOC components are present, the peculiar dependence of the Majorana polarization in Eq.~(\ref{majpol}) on the $\cos(2\phi_1)$ term indicates that the Majorana polarization vector should exhibit a spatial precession in Majorana space; we have verified numerically that this is indeed the case.

\begin{figure}[h]
\centering
\includegraphics[width=0.38\textwidth]{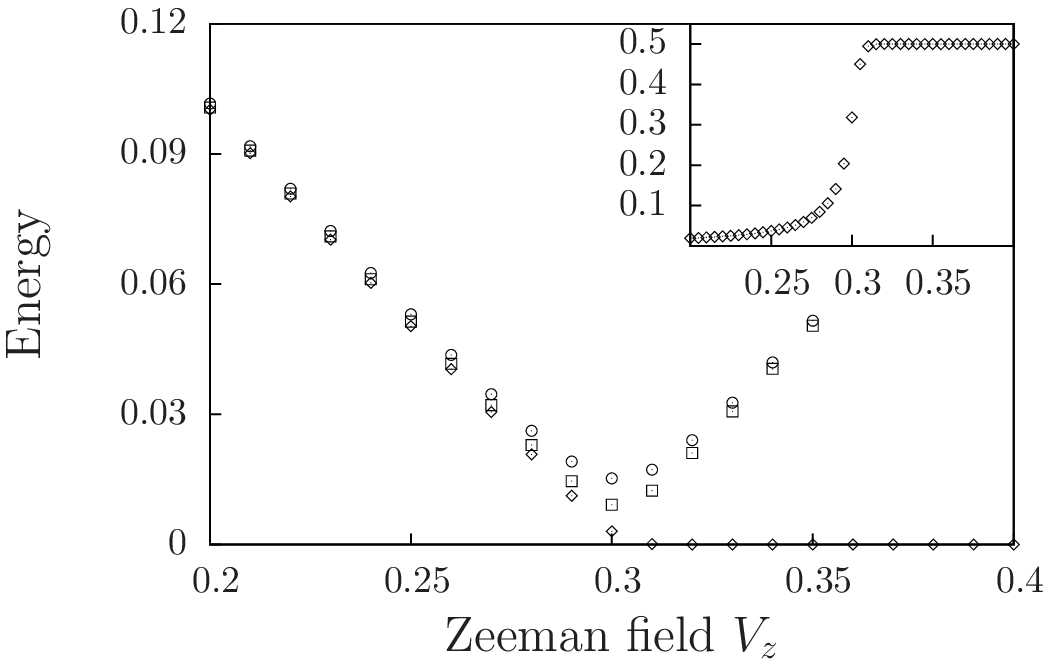}
\includegraphics[width=0.45\textwidth]{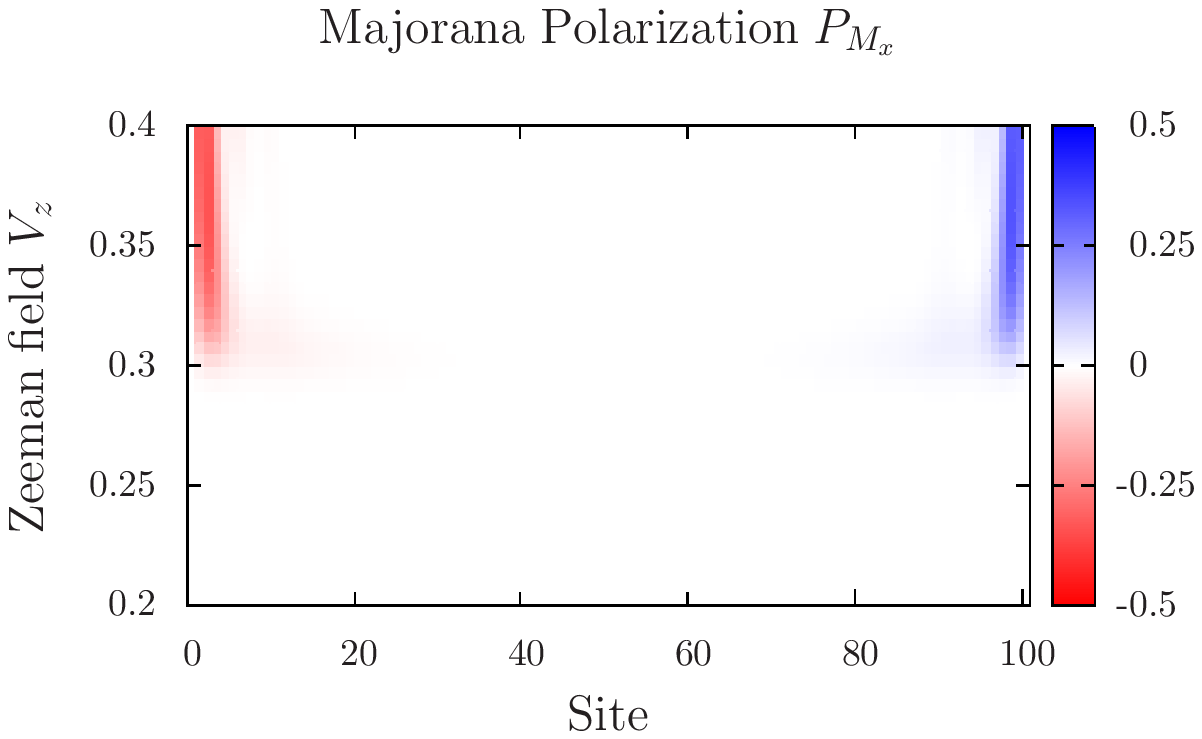}
\caption{\small First panel: lowest-energy eigenvalues and the half-wire integral of the Majorana polarization (inset) as a function of $V_z$. Second panel: Majorana polarization of the lowest-energy state as a function of position and $V_z$. Parameters: $\Delta=0.3$, $\mu=0$, $\beta=0$, and $\alpha=0.2$}
\label{fig:majv}
\end{figure}

We examine if Majorana polarization is a good order parameter to characterize the topological transition. This is done by varying one of the parameters ($\Delta, V_z, \mu$) to drive the system in a trivial phase. In Fig. (\ref{fig:majv}), we vary $V_z$, and indeed we see that the system becomes trivially gapped (no Majorana bound states) for $V_z\leq\Delta$.  The inset describes the dependence of the half-wire integral of the Majorana polarization for one of the lowest-energy states as a function of $V_z$ (an integral of $0.5$ is equivalent to a ``full'' Majorana state). The Majorana polarization decreases smoothly to zero below the critical value of $V_z$. The same phenomenon can be observed in the second panel, where we plot the spatial distribution of the Majorana polarization as a function of $V_z$. We have noted that the transition becomes sharper when increasing the size of the system. The same qualitative features are obtained when $\Delta$ and $\mu$ are varied across the topological transition (the dependence on $\mu$ is presented in SM). This shows that the Majorana polarization (and density) is a good local order parameter for the topological transition at $V_z^2=\Delta^2+\mu^2$. This suggests that the Majorana polarization can be used to investigate disordered wires\cite{disorder}, and indeed, as shown in SM, in the presence of disorder it exhibits interesting features such as a weak polarization of the low energy bulk states\cite{longpaper}. Moreover, the spin and Majorana polarization exhibit similar spatial structure, even in the presence of disorder.


{\em Conclusion}--To summarize, we have found that the Majorana end states are oppositely spin-polarized in the transverse spin-plane, and the direction of polarization depends on the relative weight of the Rashba and Dresselhaus SOC in the wire. Moreover, we have proposed a new wave-function-based measure of the Majorana character of a system, which we denote Majorana polarization. We have seen that this quantity is related to the electronic spin polarization and we have proposed to test the Majorana character of a 1D system using spin-polarized STM measurements. While the density of states measurements can only give information about the existence of a localized state at a given energy, without telling anything about its Majorana character, such a spin-polarized measurement can make the difference between a Majorana excitation or a non-topological localized state.

{\bf Aknowledgements}\\
\noindent
We would like to thank J. Alicea, S. Chakravarty, M. Franz, L. Glazman, A. Kitaev, L.-K. Lim, F. Pi\'echon, S. Raghu, G. Refael for interesting discussions. C.B. has been funded by the ERC Starting Independent Researcher Grant NANO-GRAPHENE 256965, in part by the National Science Foundation under Grant No. 1066293, and acknowledges the hospitality of the Aspen Center for Physics where part of this work has been done.

\begin{figure}[t]
\centering
\includegraphics[width=0.38\textwidth]{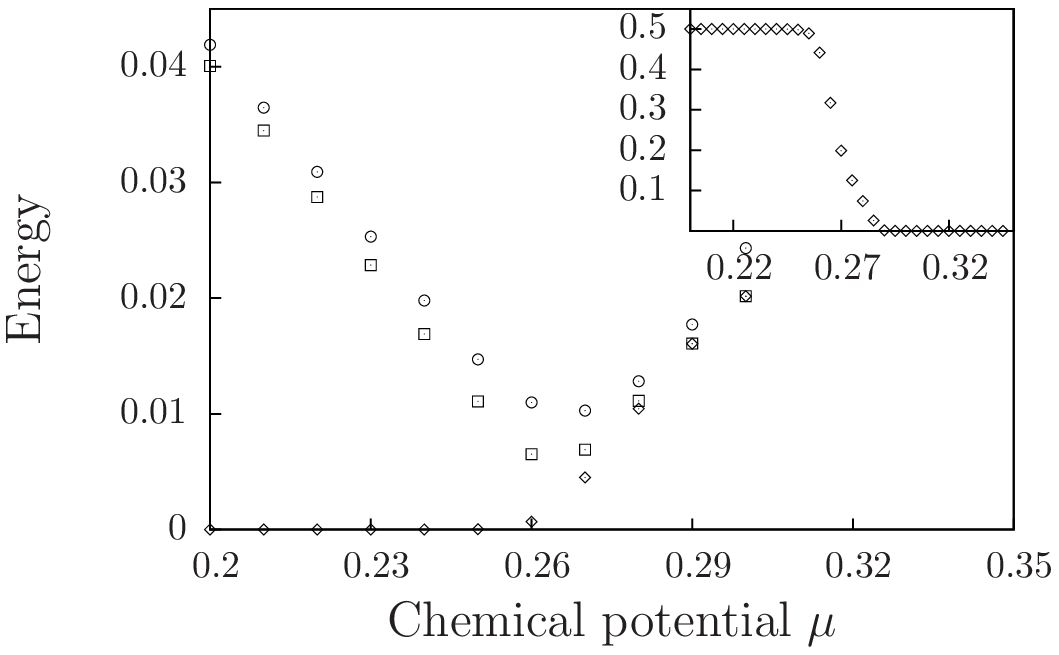}
\includegraphics[width=0.45\textwidth]{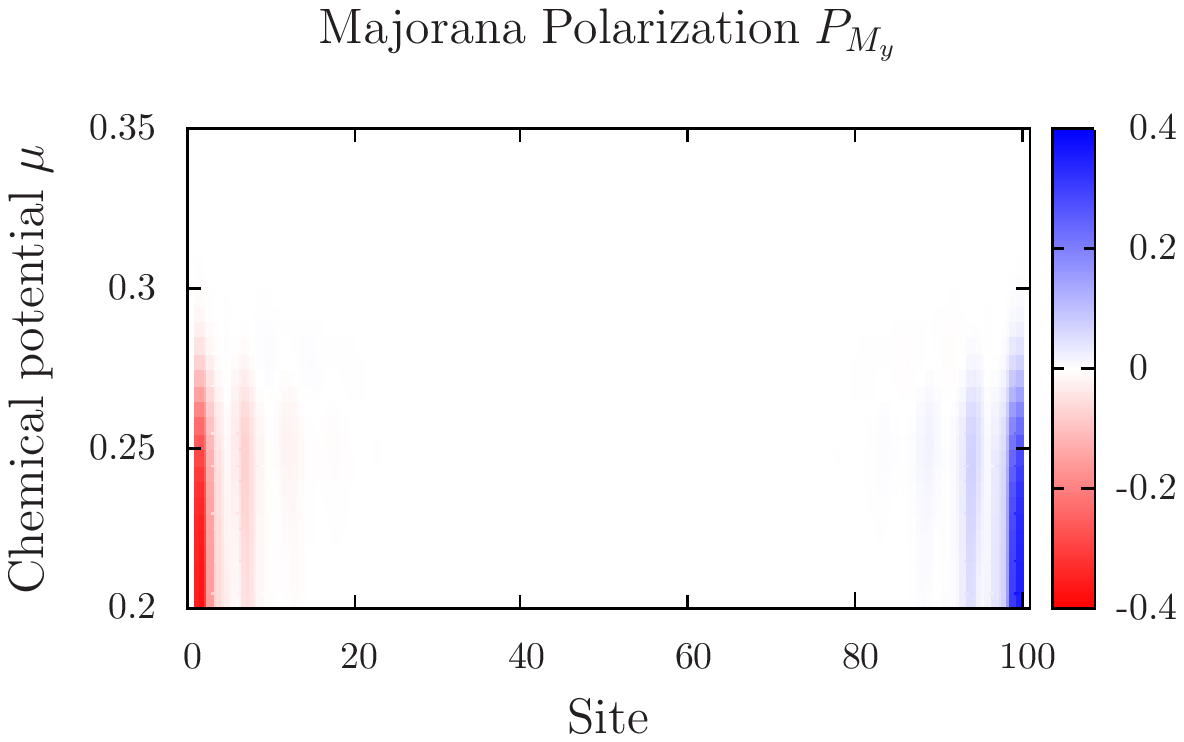}
\caption{\small In the first panel, the lowest-energy eigenvalues and the half-wire Majorana polarization integral (inset) are plotted as a function of $\mu$. In the second panel the Majorana polarization of the lowest-energy state is plotted as a function of position and $\mu$. The parameters considered are $\Delta=0.3$, $V_z=0.4$, $\alpha=0.2$, and $\beta=0$.}
\label{fig:majmu}
\end{figure}

\begin{figure}[h]
\centering
\includegraphics[width=0.4\textwidth]{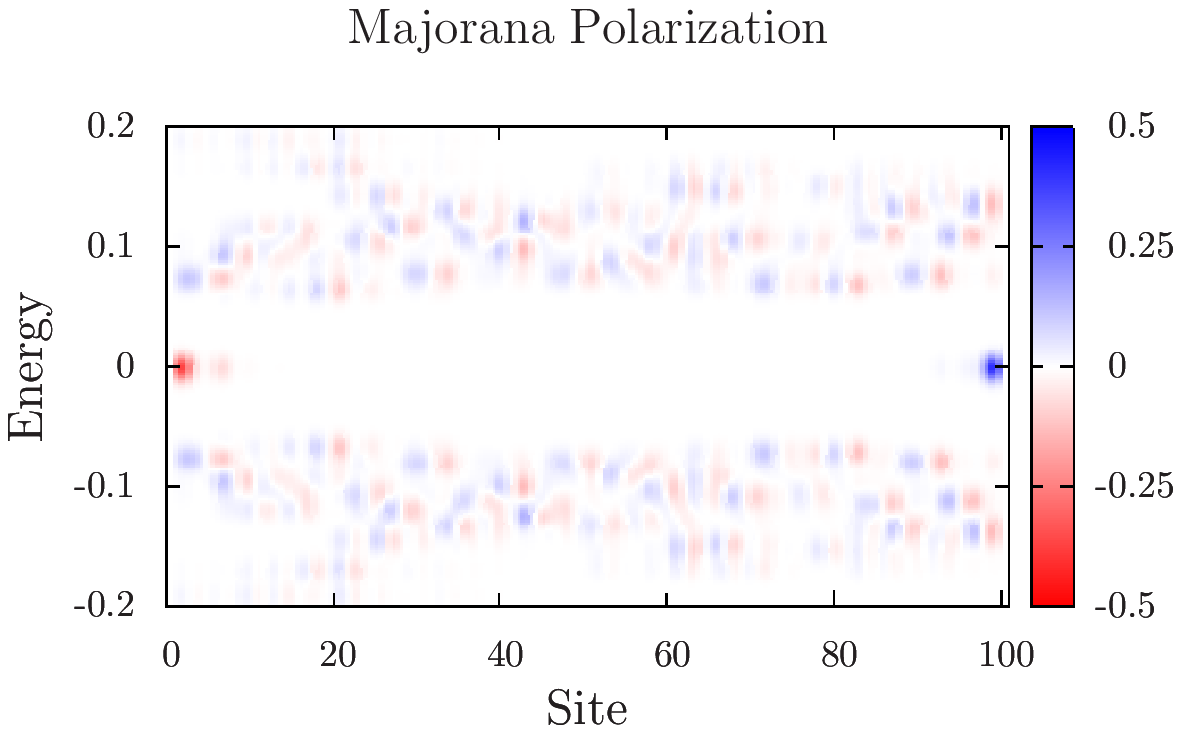}\\
\includegraphics[width=0.4\textwidth]{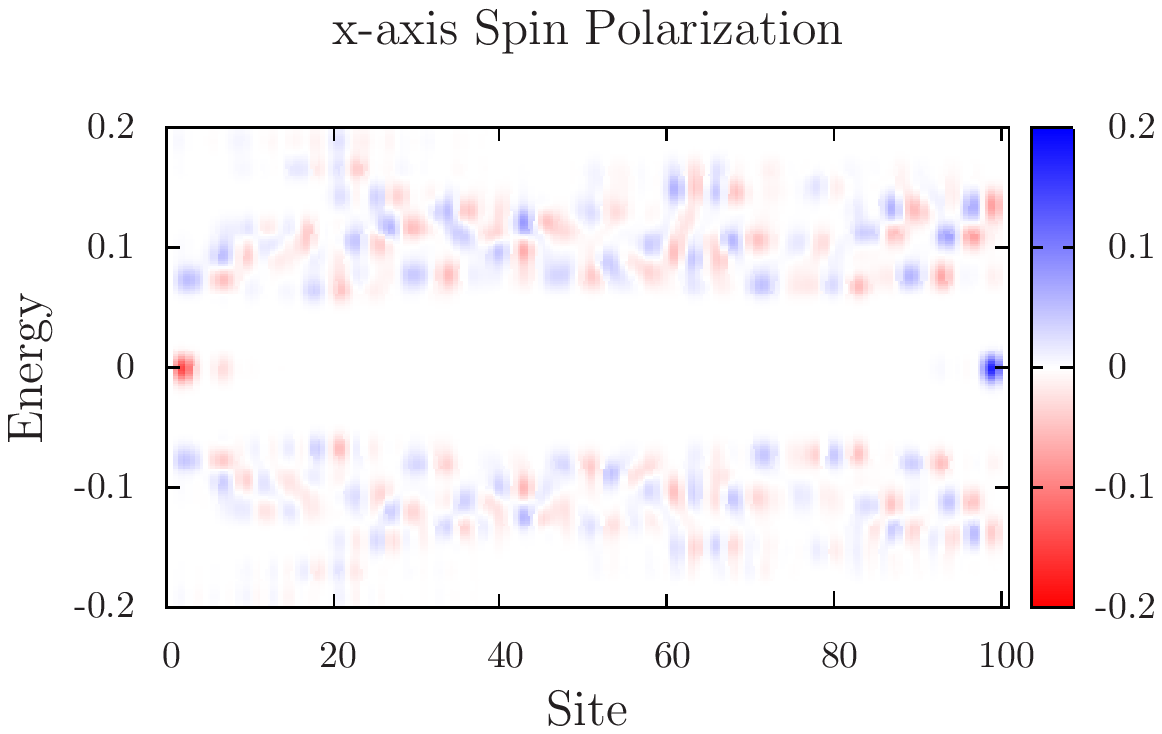}
\caption{\small The dependence of the Majorana polarization and $x$-spin-density on energy and position for a weak-disorder $w=0.5$. $\Delta=0.3$, $V_z=0.4$,
$\mu=0$, $\alpha=0.2$, and $\beta=0$}
\label{fig:w0p51}
\end{figure}

\begin{center}
\bf Supplementary Material
\end{center}
The Majorana polarization is a good order parameter to characterize the topological transition. This can be seen as well when varying the chemical potential $\mu$ in Fig.~(\ref{fig:majmu}).
We analyze also the effects of disorder on the Majorana polarization.
For example, we consider an on-site potential $W_j$ which takes random values with a uniform distribution in the interval $[-\frac{w}{2},\frac{w}{2}]$. A thorough analysis of various types of disorder (phase slips, correlated disorder, disorder in the order parameter $\Delta$ and the hopping parameter $t$, etc.) will be presented elsewhere \cite{longpaper}. For a disorder strength of $w=0.5$, when the Dresselhaus SOC $\beta$ is absent, the Majorana polarization $P_{M_x}$ still behaves qualitatively similar to the $x$-axis spin polarization, as described in Fig.~(\ref{fig:w0p51}) ($y$-polarizations remain zero). Note that, for this disorder strength, Majorana states can still be observed on the edges. Moreover the bulk states near the gap become slightly Majorana- and spin-polarized (however they are not full Majorana states).

\begin{figure}[H]
\centering
\includegraphics[width=.4\textwidth]{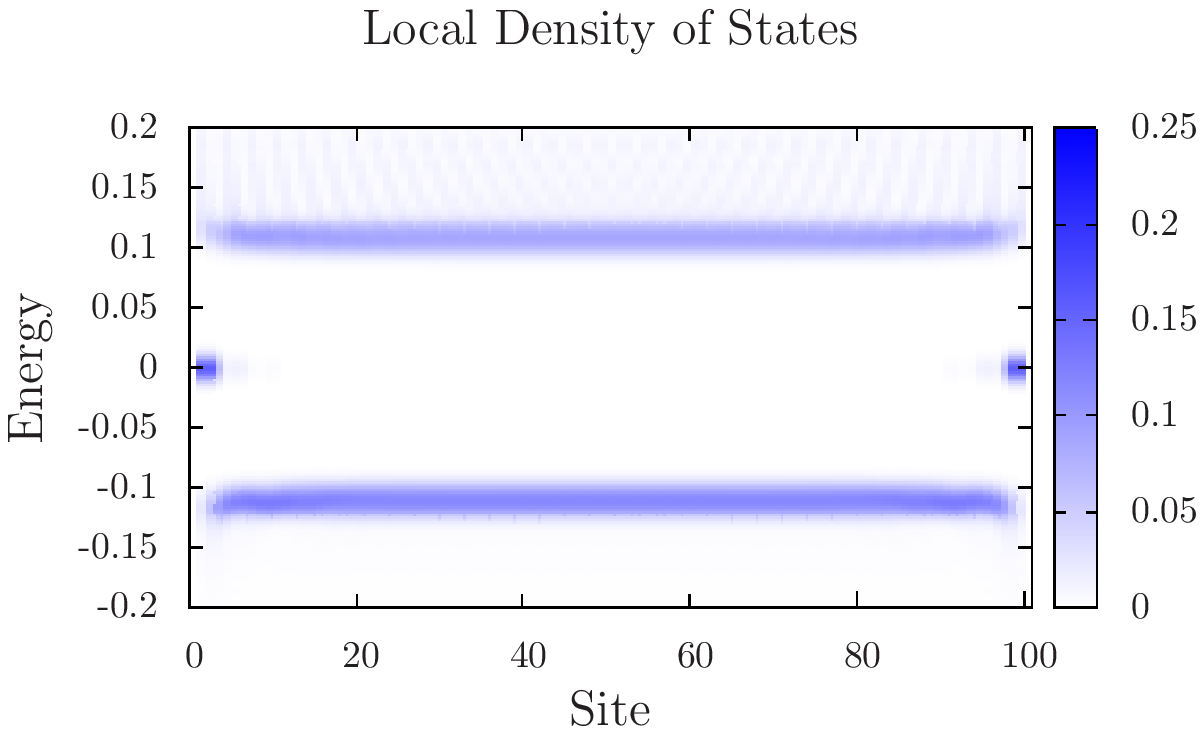}
\vspace{-0.1in}
\includegraphics[width=.4\textwidth]{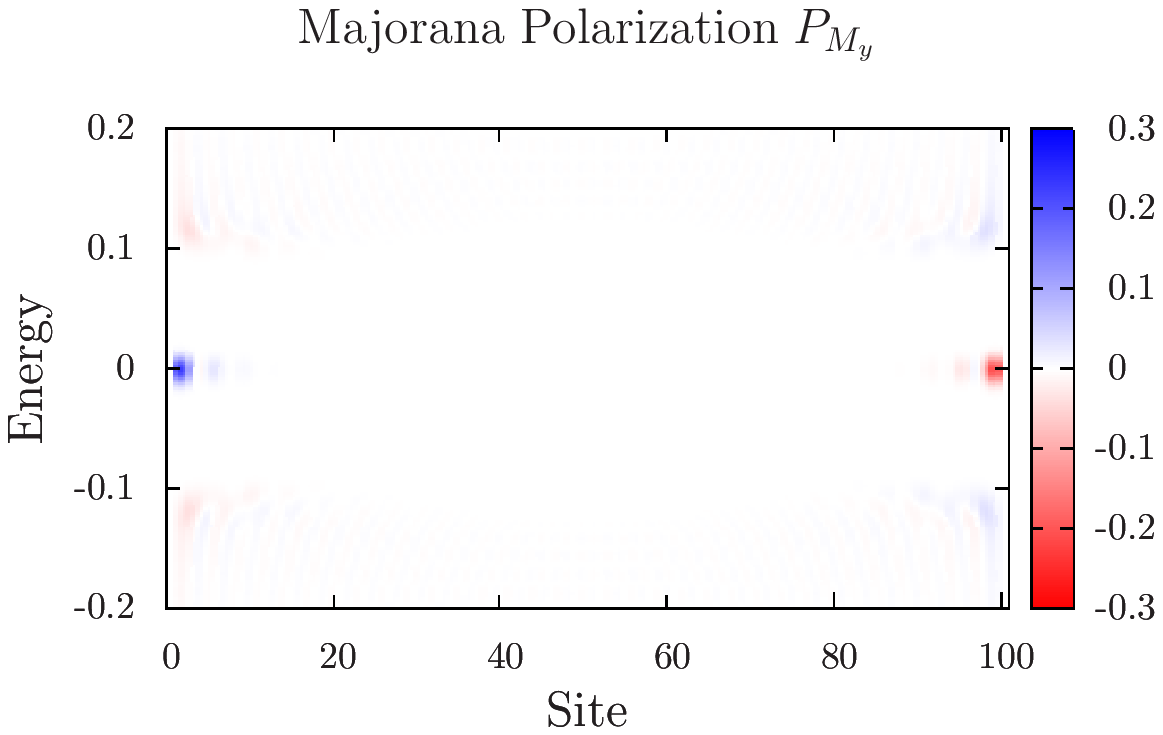}
\caption{\small Density of states and Majorana polarization $P_{M_y}$ at $\Delta=0.3$, $V_z=0.4$,
$\mu=0$, $\alpha=0$, and $\beta=0.2$.}
\label{fig:beta}
\end{figure}

Finally, we consider the presence of a Dresselhaus SOC, when the Rashba SOC is absent. We find that Majorana fermions are present, as illustrated in Fig.~(\ref{fig:beta}). The local density of states reveals zero energy edge states, and a plot of the $P_{M_y}$ identifies them as Majorana states. Here the transverse spin polarization has only a non-zero component along the $y$-axis, that, as expected, is proportional to $P_{M_y}$.\\
\end{document}